\documentclass[10pt,letterpaper,english]{article}
\usepackage[T1]{fontenc}
\usepackage[latin1]{inputenc}
\usepackage{amsmath}
\usepackage{graphicx}
\usepackage{amssymb}
\usepackage{esint}
\makeatletter

\makeatletter
\usepackage{amscd}

\usepackage{amsfonts}

\usepackage{babel}\makeatother

\makeatother

\usepackage{babel}

\begin{document}
\newtheorem{proposition}{Proposition}[section] \newtheorem{definition}{Definition}[section]
\newtheorem{corollary}{Corollary}[section] \newtheorem{lemma}{Lemma}[section]
\newtheorem{theorem}{Theorem}[section] \newtheorem{example}{Example}[section]

\title{\textbf{On the Solution of the Multi-asset Black-Scholes
		model: Correlations, Eigenvalues and Geometry} }

\author{M. Contreras$^{\dagger}$, A. Llanquihuén\thanks{Facultad de Ciencias Exactas, Universidad Andres Bello, Chile} \ and M.Villena\thanks{Facultad de Ingeniería y Ciencias, Universidad Adolfo Ibáñez, Chile}}

\maketitle 

\begin{abstract}

In this paper, we study the multi-asset Black-Scholes model in terms of the importance that the correlation parameter space (equivalent to an $N$ dimensional hypercube) has in the solution of the pricing problem. We show that inside of this hypercube there is a surface, called the Kummer surface $\Sigma_K$, where the determinant of the correlation matrix $\rho$ is zero, so the usual formula for the propagator of the $N$ asset Black-Scholes equation is no longer valid. Worse than that, in some regions outside this surface, the determinant of $\rho$ becomes negative, so the usual propagator becomes complex and divergent. Thus the option pricing model is not well defined for these regions outside $\Sigma_K$. On the Kummer surface instead, the rank of the $\rho$ matrix is a variable number. By using the Wei-Norman theorem, we compute the propagator over the variable rank surface $\Sigma_K$ for the general $N$ asset case. We also study in detail the three assets case and its implied geometry along the Kummer surface.	
\end{abstract}

\section{Introduction}

Since the seminal work of Black, Scholes and Merton on option pricing, see \cite{1} and \cite{2}, an important research agenda has been developed on the subject. This research has mainly centered in extending the basic Black and Scholes model to well known empirical regularities, with the hope of improving the predicting power for the famous formula, see for example \cite{4}, \cite{5}, \cite{8}, \cite{9}. An interesting extension has been the modeling of many underlying assets, which has been called the multi-asset Black-Scholes model \cite{4}, \cite{11}. In this case, the option price satisfies a diffusion equation considering many related assets. 
The first work addressing this problem in the literature was Margrabe (1978), see \cite{18}. The Margrabe formula considered an exchange option, which gives its owner the right, but not the obligation, to exchange $b$ units of one asset into a unit of another asset at a specific point in time. Specifically, Margrabe derived a closed-form expression for the option by taking one of the underlying assets as a numeraire and then applying the Black and Scholes standard formulation. Later Stulz \cite{19} found analytical formulae for European put and call options on the minimum or the maximum of two risky assets. In this particular case, the solution is expressed in terms of bivariate cumulative standard normal distributions, and when the strike price of the option is zero the value reduces to the Margrabe pricing. Other interesting papers that follow in this literature are \cite{20}, \cite{21}, \cite{22}, \cite{23}, \cite{24}, \cite{25}. The numerical implementation of the solution of the multi-asset Black-Scholes model is increasingly difficult for models with more that three assets, see for instance \cite{12}, \cite{13}, \cite{14}. One important point, that has been missed in the literature, is that in all of the multi-asset Black-Scholes models mentioned above, the relationship between assets is modeled by their correlations, and hence it is implicitly assumed that a well behaved multivariate Gaussian distribution must exist in order to have a valid solution.  \\
In this paper, we study the multi-asset Black-Scholes model in terms of the importance that the correlation parameter space (which is equivalent to an $N$ dimensional hypercube) has in the solution of the option pricing problem. We show that inside of this hypercube there is a surface, called the Kummer surface $\Sigma_{K}$ \cite{35}, \cite{36}, \cite{37}, \cite{38}, where the determinant of the correlation matrix $\rho$ is zero, so over $\Sigma_K$ the usual formula for the propagator of the $N$ asset Black-Scholes equation is no longer valid. Worse than that, outside this surface, the are points where the determinant of $\rho$ becomes negative, so the usual propagator becomes complex and divergent. Thus the option pricing model is not well defined for some regions outside $\Sigma_{K}$. On $\Sigma_{K}$ the rank of $\rho$ matrix is a variable number, depending on which sector of the Kummer surface the correlation parameters are lying. By using the Wei-Norman theorem \cite{39}, \cite{40}, \cite{41}, \cite{42}, we found the propagator along the Kummer surface $\Sigma_K$, for the $N$ assets case. Our expression is valid  whatever be the value of the $\rho$ matrix rank over $\Sigma_K$. \\ \\
This paper is organized as follows. Section 2 describes the traditional multi-asset Black-Scholes model. In section 3, the problem is formulated as a $N$ dimensional diffusion equation. In section 4, the implied geometry of the correlation matrix space is analyzed, specially when its determinant is zero, which coincides with a Kummer surface in algebraic geometry. The Kummer surface and its geometry is reviewed for the particular case of three assets in section 4.1. In section 5, by using the Wei-Norman theorem  the propagator over the variable rank surface $\Sigma_{K}$ for a general $N$ asset case is computed. Finally, some conclusions and future research are presented in section 6.

\section{The multi-asset Black-Scholes model}

Consider a portfolio consisting of one option and $N$ underlying assets. Let $S_i$ be the price processes for the assets; $i=1...N$ where each asset satisfies the usual dynamic
\begin{equation}\label{Prices}\mathit{dS}_{i}=\alpha _{i}S_{i}\mathit{d\tau}+\sigma_{i}S_{i}\mathit{dW}_{i}
\end{equation}
\noindent $i=1...N$ and the $N$ Wiener processes $W_i$ are correlated according to
\begin{equation}\mathit{dW}_{i}\mathit{dW}_{j}= \rho_{\mathit{ij}}\mathit{d\tau} \end{equation}
\noindent where $\rho$ is the symmetric matrix
\begin{equation} \label{rhomatrixNxN}
\rho = \left(
	\begin{matrix} 1 & \rho_{12} & \rho_{13} & \rho_{14}\cdots &\rho_{1N}\\
		\rho_{12} & 1 & \rho_{23} & \rho_{24}\cdots &\rho_{2N}\\
		\vdots &\vdots &\vdots&&\vdots \\
		\rho_{1N} & \rho_{2N} & \rho_{3N} & \rho_{4N}\cdots & 1 \\
	\end{matrix}\right)
\end{equation}
\noindent so we have 
\begin{equation}\mathit{dS}_{i}\mathit{dS}_{j}=\sigma _{i}\sigma _{j}S_{i}S_{j}\rho_{\mathit{ij}}\mathit{d\tau}
\end{equation} \\
If the price process for the option is $\Pi=\Pi (S_{1},S_{2,}...S_{n},\tau)$, the value $V$ of the portfolio is given by
\begin{equation}V=\Pi -\sum _{i}\Delta _{i}S_{i}\end{equation}
\noindent where $\Delta _{i}$ are the shares of each asset in the portfolio. The self-financing portfolio condition ensures that
\begin{equation}\label{Selffinancing}\mathit{dV}=d\Pi -\sum _{i}\Delta _{i}\mathit{dS}_{i}\end{equation}
\noindent and applying It\^{o} Lemma for $\Pi$ one gets
\begin{equation}\label{Portfolio}\mathit{dV}=\left(\frac{\partial \Pi }{\partial \tau} d \tau + \sum _{i}  \frac{\partial \Pi
	}{\partial S_{i}}\mathit{dS}_{i}+ \sum _{i,j} \frac{1}{2}\frac{\partial ^{2}\Pi
}{\partial S_{i}\partial
S_{j}}\mathit{dS}_{i}\mathit{dS}_{j}\right)-\sum _{i}\Delta
\mathit{dS}_{i}\end{equation}
According to \cite{4}, for a free arbitrage set of $N$ assets, the return of the portfolio is 
\begin{equation}
\label{Arbitrage}\mathit{dV}=\mathit{rVd \tau}
\end{equation} 
\noindent and from equations (\ref{Portfolio}) and (\ref{Arbitrage}) one has
\begin{equation}\begin{gathered}\frac{\partial \Pi }{\partial
			\tau}\mathit{d\tau}+\sum _{i}\frac{\partial \Pi }{\partial S_{i}}\left(\alpha
		_{i}S_{i}\mathit{d\tau}+\sigma_{i}S_{i}\mathit{dW}_{i}\right)+\sum _{i,j}\frac{1}{2}\frac{\partial ^{2}\Pi
		}{\partial S_{i}\partial S_{j}}\sigma _{i}\sigma
		_{j}S_{i}S_{j}\rho _{\mathit{ij}}\mathit{d\tau}-\sum _{i}\Delta
		_{i}\left(\alpha _{i}S_{i}\mathit{d\tau}+\sigma
		_{i}S_{i}\mathit{dW}_{i}\right)\\=r\left(\Pi -\sum _{i}\Delta
		_{i}S_{i}\right)\mathit{d\tau} \end{gathered}\end{equation}
\noindent Collecting $\mathit{d\tau}$ and $\mathit{dW_i}$ terms in the above equation one gets:
\begin{equation}\label{dt}\frac{\partial \Pi }{\partial \tau}+\sum _{i}{}\frac{\partial \Pi
	}{\partial S_{i}}\alpha _{i}S_{i}+\sum
	_{i,j}{}\frac{1}{2}\frac{\partial ^{2}\Pi }{\partial S_{i}\partial
		S_{j}}\sigma _{i}\sigma _{j}S_{i}S_{j}\rho _{\mathit{ij}}-\sum
	_{i}\Delta _{i}\alpha _{i}S_{i}-r\left(\Pi -\sum _{j}\Delta
	_{j}S_{j}\right)=0\end{equation}
\noindent and
\begin{equation}\label{dW}\sum _{i}{}\left[\frac{\partial \Pi }{\partial S_{i}}\sigma
	_{i}S_{i}-\Delta_{i}\sigma_{i} S_{i}\right]\mathit{dW}_{i}=0\end{equation}
\noindent From equation (\ref{dW}), and given the independence of the $W_i$ , we can say that for $i=1...N$
\begin{equation}\frac{\partial \Pi }{\partial S_{i}}\sigma _{i}S_{i}-\Delta _{i}\sigma
_{i}S_{i} = 0 
\end{equation}
or equivalently
\begin{equation}\label{dWi}
\Delta_{i} = \frac{\partial \Pi }{\partial S_{i}} 
\end{equation}
so one arrives at the multi-asset Black-Scholes equation
\begin{equation}\label{Main}
	\begin{array}{rcl}
	\displaystyle \frac{\partial \Pi }{\partial \tau}+\sum_{i,j}{\frac{1}{2}}\frac{\partial ^{2}\Pi }{\partial S_{i}\partial
		S_{j}}\sigma _{i}\sigma _{j}S_{i}S_{j}\rho _{\mathit{ij}}+ r \left(\sum_{j} S_j \frac{\partial \Pi}{\partial S_j}-\Pi\right) & = & 0\\
	\end{array}
\end{equation}
which must be integrated with the final condition
$$
\Pi(\vec{S},T)  =  \Phi(\vec{S})
$$
for constant $r$, $\alpha_i$, $\sigma_i$
and a simple contingent claim $\Phi$.

\section{The multi-asset Black-Scholes equation as a $N$ dimensional diffusion equation}

Here, some transformations are developed, which maps the multi-asset option pricing equation in a more simpler diffusion equation. If one makes the change of variables
\begin{equation} \label{logaS}
	x_i =\ln(S_i)-(r-\frac{1}{2} \sigma_i^2) \tau
\end{equation}
in (\ref{Main}), one can map this equation to
$$
\displaystyle \frac{\partial \Pi }{\partial \tau}+{\frac{1}{2}} \sum
_{i,j} \sigma_i \sigma_j \rho_{ij} \frac{\partial ^{2}\Pi }{\partial x_{i}\partial
	x_{j}} - r \Pi = 0
$$
At least if one defines $\Psi$ as
\begin{equation} \label{PiPsi}
\Pi(\vec{x}, \tau) =e^{-r(T-\tau)} \Psi(\vec{x},\tau)
\end{equation}
then $\Psi$ satisfies the equation
$$
\displaystyle \frac{\partial \Psi }{\partial \tau}+{\frac{1}{2}} \sum
_{i,j} \sigma_i \sigma_j \rho_{ij} \frac{\partial ^{2}\Psi }{\partial x_{i}\partial
	x_{j}} = 0
$$
Now, by defining the variables
\begin{equation} \label{chisigma}
\chi_i = \frac{x_i}{\sigma_i}
\end{equation}
the above equation can be written as
$$
\displaystyle \frac{\partial \Psi }{\partial \tau}+{\frac{1}{2}} \sum
_{i,j} \rho_{i j}\frac{\partial ^{2}\Psi }{\partial \chi_{i}\partial \chi_{j}} = 0
$$
And finally, by defining the forward time coordinate
\begin{equation} \label{forwardtime}
t = T-\tau
\end{equation}
one arrives at 
\begin{equation} \label{EuclideanSE}
\frac{\partial \Psi }{\partial t} = {\frac{1}{2}} \sum
_{i,j} \rho_{i j}\frac{\partial ^{2}\Psi }{\partial \chi_{i}\partial \chi_{j}} 
\end{equation}
Now performing the transformation 
\begin{equation} \label{zetachi}
\vec{\zeta} = U^{-1} \vec{\chi} 
\end{equation}
one can change the $\chi_k$ variables to the  $\zeta_k$ coordinates that diagonalizes the $\rho$ matrix
\begin{equation}
D = U^{-1} \rho \ U
\end{equation}
where
\begin{equation}
D = diag(\lambda_1,\lambda_2, ... \lambda_N)
\end{equation}	
and $U$ is the change basis matrix, with $U^{-1}=U^t$,  $\det(U)=1$. The explicit form of the $U$ matrix in terms of the $(x,y,z)$ variables is very complex and we do not write it explicitly. In this diagonal coordinate system, the diffusion equation read finally
\begin{equation} \label{difussiondiag}
\frac{\partial \Psi }{\partial t} = \frac{1}{2} \sum_{i=1}^N \lambda_{i} \frac{\partial^{2}\Psi}{\partial \zeta_{i}^2} 
\end{equation}
Now we study this equation in terms of the behavior of the eigenvalues $\lambda_i$.  

\section{The geometry of the $\rho$ matrix}
The $\rho$ matrix in (\ref{rhomatrixNxN}) can be characterized completely for the $M =\frac{N(N-1)}{2}$ dimensional vector
\begin{equation}
\vec{r}= (\rho_{12}, \rho_{13}, \rho_{14}, ... \ , \rho_{(N-1) N})   \ \ \ \ \ \ \ -1 \leq \rho_{ij}\leq 1
\end{equation}
which lies inside of an $M$ dimensional hypercube centering in the origin and of length 2. Thus, the $\rho$ matrix is a function of $\vec{r}$: $ \rho = \rho(\vec{r}) $. Note that, for some point $\vec{r}$ inside of the hypercube, the determinant of the $\rho$ matrix vanishes. For example, for the vertex
\begin{equation} \label{vertex1}
\vec{r}= (1, 1, 1, ... \, 1) \ \ \ \ \  \Rightarrow  \ \ \ \ \ \det(\rho) = 0
\end{equation}
In fact, exists a whole surface inside the hypercube, where the determinant of $\rho$ vanishes. This surface, is called the Kummer surface $\Sigma_K$ in algebraic geometry \cite{35}, \cite{36}, \cite{37}, \cite{38}, is defined by the equation
\begin{equation} \label{deternull}
\vec{r} \in \Sigma_0 \ \ \ \Leftrightarrow \ \ \ \det \left(
\begin{matrix} 1 & \rho_{12} & \rho_{13} & \rho_{14}\cdots &\rho_{1N}\\
\rho_{12} & 1 & \rho_{23} & \rho_{24}\cdots &\rho_{2N}\\
\vdots &\vdots &\vdots&&\vdots \\
\rho_{1N} & \rho_{2N} & \rho_{3N} & \rho_{4N}\cdots & 1 \\
\end{matrix}\right) = \det \rho(\vec{r})=0
\end{equation}
In fact, one can think of the hypercube as the disjoint union of the subset of point or surfaces $\Sigma_C$ of constant $C$ determinant value:
\begin{equation} \label{deterconstant}
\vec{r} \in \Sigma_C \ \ \ \Leftrightarrow \ \ \ \det  \left(
\begin{matrix} 1 & \rho_{12} & \rho_{13} & \rho_{14}\cdots &\rho_{1N}\\
\rho_{12} & 1 & \rho_{23} & \rho_{24}\cdots &\rho_{2N}\\
\vdots &\vdots &\vdots&&\vdots \\
\rho_{1N} & \rho_{2N} & \rho_{3N} & \rho_{4N}\cdots & 1 \\
\end{matrix}\right) = \det \rho(\vec{r}) = C
\end{equation}
Let $\vec{r}$ an arbitrary vector in $\mathbb{R}^M$ and let $\phi(\vec{r})$ the determinant of $\rho$ in each point, that is $\phi(\vec{r})=\det(\rho(\vec{r}))$. Note that $\phi(\vec{r})$ is a polynomial function in terms of the $\vec{r}$ coordinates. \\
The vector $\vec{\eta}$ given by the $M$ dimensional gradient $\vec{\eta} = \nabla_{\vec{r}} \ \phi(\vec{r}) $ \ is perpendicular to the level surfaces $\Sigma_C$ and gives the direction for greater growth of the function $\phi(\vec{r})$. Note also that the components of this vector are also polynomial functions of the $\vec{r}$ coordinates, so  $\vec{\eta}=\vec{\eta}(\vec{r}) $ is a continuous vector function. \\ \\
Consider now a point $\vec{r_0} \in \Sigma_K$ , that is, $\phi(\vec{r_0})=0$. As $\phi$ and $\vec{\eta}$ are continuous, there is a neighbor of $\vec{r_0}$ on $\Sigma_K$, such that for $\epsilon > 0$ the vector $\vec{r}_{+} = \vec{r_0}+\epsilon \vec{\eta} \ \in \Sigma_C$ with $C > 0$, whereas the vector  $\vec{r}_{-} = \vec{r_0}-\epsilon \vec{\eta} \ \in \Sigma_C$ with $C < 0$, \ due to the $\phi$ function growths along the $\vec{\eta}$ direction. Thus, the Kummer surface $\Sigma_K$ separates spacial regions with positive $\rho$ determinant from that with negative $\rho$ determinant. \\ \\
In its diagonal form, equation (\ref{deternull}) is
\begin{equation} \label{deterdiagonal}
\vec{r} \in \Sigma_K \ \ \ \Leftrightarrow \ \ \ \det  \left(
\begin{matrix} \lambda_1 & 0 & 0 & 0 \cdots & 0 \\
0  & \lambda_2 & 0 & 0 \cdots & 0  \\
\vdots &\vdots &\vdots&&\vdots \\
0  &  0 & 0 & 0 \cdots & \lambda_N \\
\end{matrix}\right) = 0
\end{equation}
where the $\lambda_i = \lambda_i(\vec{r})$, that is
\begin{equation} \label{deterdiagonal2}
\vec{r} \in \Sigma_K \ \ \ \Leftrightarrow  \ \ \  \phi(\vec{r})= \lambda_1(\vec{r}) \lambda_2(\vec{r}) \cdots \lambda_n(\vec{r}) = 0
\end{equation}
Note that equation (\ref{deterdiagonal2}) implies that there is at least one eigenvalue that is zero over all the Kummer surface. But on $\Sigma_K$ other eigenvalues can also become null. Thus, the Kummer surface is a variable $\rho$ rank surface. \\ \\
As $\phi(\vec{r})$ is equal to $ \lambda_1(\vec{r}) \lambda_2(\vec{r}) \cdots \lambda_n(\vec{r})$, the vector $\vec{\eta}$ can be written as
\begin{equation}
\begin{matrix}
\vec{\eta}(\vec{r})=& [\nabla_{\vec{r}} \lambda_1(\vec{r})] \lambda_2(\vec{r}) \cdots \lambda_n(\vec{r}) +   \\
& \lambda_1(\vec{r}) [\nabla_{\vec{r}} \lambda_2(\vec{r})] \cdots \lambda_n(\vec{r}) +  \\
& \lambda_1(\vec{r}) \lambda_2(\vec{r}) \cdots [\nabla_{\vec{r}}  \lambda_n(\vec{r})]   \\
\end{matrix}
\end{equation}
Let say that $\lambda_1$ is the zero eigenvalue over all Kummer surface. Then over $\Sigma_K$, the vector $\vec{\eta}$ is given by
\begin{equation} \label{etasigmacero}
\vec{\eta}(\vec{r}) = [\nabla_{\vec{r}} \lambda_1(\vec{r})] \lambda_2(\vec{r}) \cdots \lambda_n(\vec{r}) 
\end{equation}
If $\Sigma_{K_n}$ is the subregion of $\Sigma_K$ over which there are $n>1$ null eigenvalues, then by (\ref{etasigmacero})
\begin{equation} \label{etanull}
\vec{\eta}(\vec{r}) = 0 \ \ \ \  \forall \ \vec{r} \in \Sigma_{K_n}
\end{equation}
Thus higher order rank subregions $\Sigma_{K_n}$ of the Kummer surface are characterized by the fact that the $\vec{\eta}$ vector vanishes on them. \\ \\
Consider now, the origin $\vec{r_O}=(0,0, \cdots, 0)$ where $\phi(\vec{r_O})=1$. It is easy to show that for points $\vec{r}$ near to the origin, the function $\phi$ goes as $\phi(\vec{r}) \approx 1-\lVert \vec{r}\rVert^2$ by expanding $\phi$ in Taylor series around the origin and keeping the least order terms in the expansion. The $\vec{\eta}$ vector near the origin is then $\vec{\eta} =-2\vec{r}$ and its an inward radial vector. So near the origin, the constant determinant surfaces $\Sigma _C$ are given approximately by $M$ dimensional spheres and $\phi$ growths inward to the origin. \\ \\
Let  $\Gamma$ a curve that starts in the origin and that is normal to all $\Sigma_C$ surfaces, that is, its tangent vector is parallel to the $-\vec{\eta}$ vector in each point. Because, near the origin the vector $-\vec{\eta}$ is radial, one can reach any point of the space starting from the origin using such a curve. Moving  along $\Gamma$ in the outer direction, the $\phi$ function always decreases from its initial value 1. Thus, at some point $\vec{r_0}$ \ in $\Gamma$, the $\phi$ function vanishes. Thus means that the Kummer surface $\Sigma_K$ must contain a closed subsurface $\Sigma_0$ that enclosed the origin. Then inside of this closed subsurface $\Sigma_0$ the determinant of the $\rho$ matrix must be positive and outside $\Sigma_0$ there are points where the determinant of the correlation matrix is necessarily negative. Note that $\Sigma_0$ can be contained totally inside the hypercube or can cut it in different regions with positive or negative determinant values respectively. \\ \\
Thus, outside $\Sigma_0$ there are regions where the determinant 
\begin{equation}
\lambda_1 \lambda_2 \cdots \lambda_N < 0
\end{equation}
so at least one of the eigenvalues must be negative outside $\Sigma _0$. Inside $\Sigma_0$ however
\begin{equation}
\lambda_1 \lambda_2 \cdots \lambda_N > 0
\end{equation}
This implies that pairs of eigenvalues can be negative. But inside $\Sigma_0$ the eigenvalue cannot be negative. To prove that, consider the origin $\vec{r}_O$ where all eigenvalues  $\lambda_i = \lambda_i(\vec{r_O})$ are equal to one. When $\vec{r}$ moves outward along a curve $\Gamma$ that start at the origin, each eigenvalue  $\lambda_i = \lambda_i(\vec{r})$ will change its value from its initial positive value 1, but cannot become negative. If $\lambda_i = \lambda_i(\vec{r}) < 0$ for some points $\vec{r}$ along $\Gamma$ inside of $\Sigma_0$, then there is a point $\vec{r_0}$ where $\lambda_i=0$. This implies that the vector $\vec{r}$ would cross the surface $\Sigma_0$, but it is impossible because $\vec{r}$ is inside of $\Sigma_0$ where $\det\rho > 0$. Then inside the surface $\Sigma_0$ all eigenvalues of the correlation matrix are positive. \\ \\
In order to grasp the above ideas we study in detail the case of three assets in the next sub section.
\subsection{The geometry of the $N = 3$ assets case}
The $\rho$ matrix, for the  three assets case, is equal to
\begin{equation}
\rho = \left(
\begin{matrix} 1 & \rho_{12} & \rho_{13} \\
\rho_{12} & 1 & \rho_{23} \\
\rho_{13} & \rho_{23} & 1 \\
\end{matrix}\right)      =        
\left(\begin{matrix}
	1 & x & y \\
	x & 1 & z \\
	y & z & 1 \\
\end{matrix}\right) \label{rhomatrix}
\end{equation}
where we write the vector $\vec{r} = (\rho_{12}, \rho_{13}, \rho_{23})$ as $\vec{r}= (x, y, z)$. For this parameterization the determinant of the $\rho$ matrix is
$$
\det(\rho)=2xyz-x^2-y^2-z^2+1
$$
The constant determinant $\Sigma_C$ surfaces \ $\det(\rho(\vec{r})) = C$ \ in the interior of the hypercube are shown in figure 1, for some positive values between $0<C<1$. Instead, in figure 2, some surfaces for negative $C$ values are displayed with $-3<C<0$.
\begin{figure}[h] 
	\centering \includegraphics[scale=0.30]{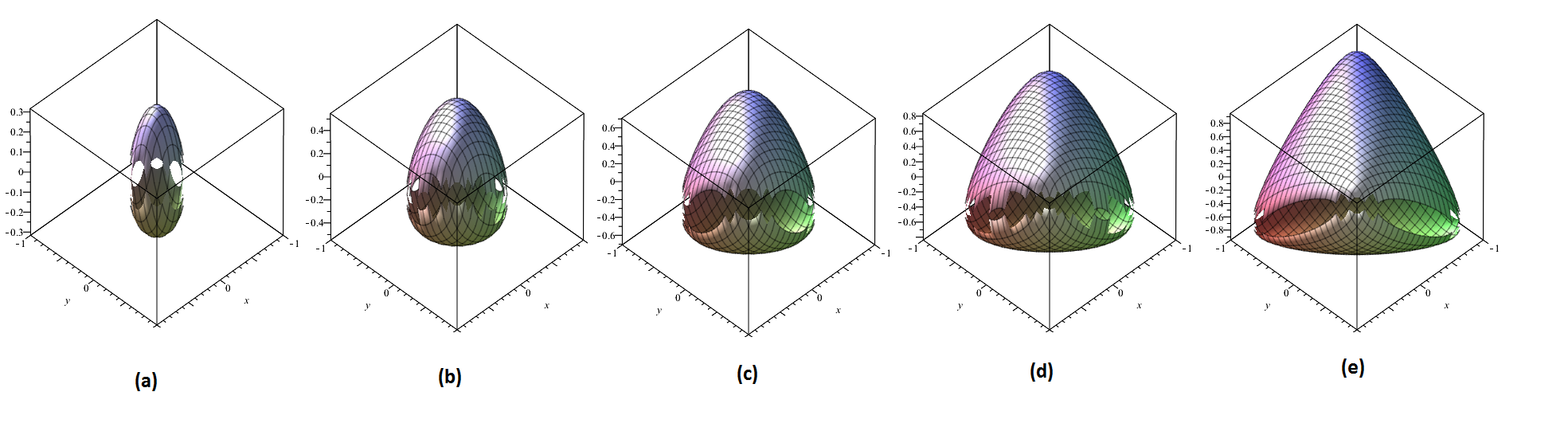} \caption{(a) $C=0.9$, (b) $C=0.7$, (c) $C=0.5$, (d) $C=0.3$, (e) $C=0.1$ } \label{positiveCsurface}
\end{figure}\\
\begin{figure}[h] 
	\centering \includegraphics[scale=0.30]{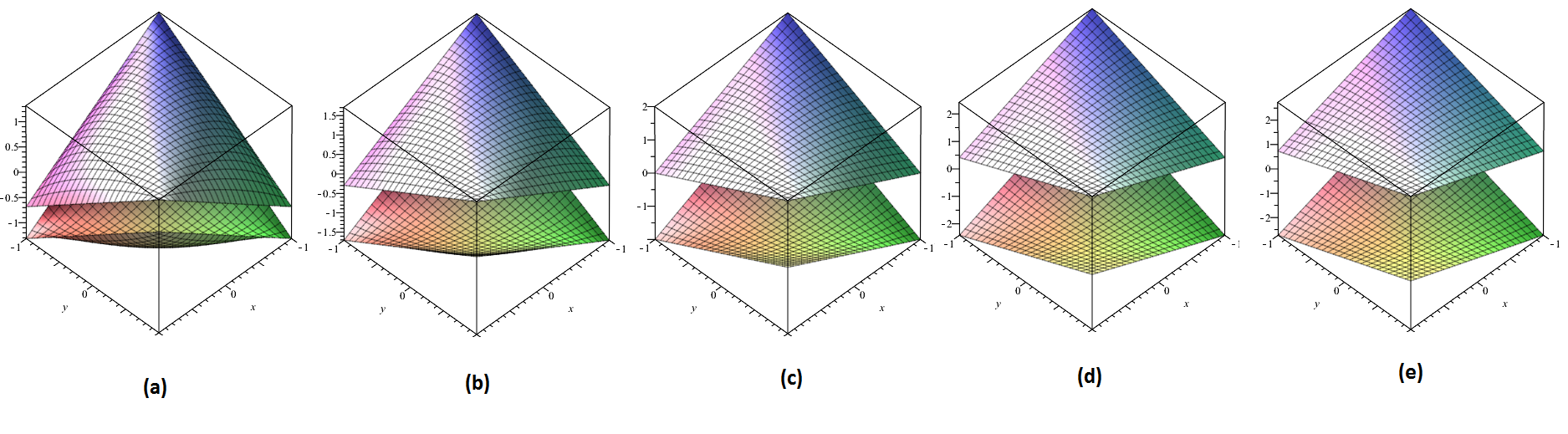} \caption{(a) $C=-0.1$, (b) $C=-0.5$, (c) $C=-1$, (d) $C=-2$, (e) $C=-3$ } \label{negativeCsurface}
\end{figure}\\
The Kummer $\Sigma_K$ surface is given by the condition $\det \rho(\vec{r})=0$, that is
\begin{equation} \label{nullsurfacedefi}
2xyz-x^2-y^2-z^2+1=0
\end{equation}
From (\ref{nullsurfacedefi}) one found that the Kummer $\Sigma_0$ subsurface inside the hypercube is given by the parametric equations
\begin{equation}
z=z^{\pm}(x,y)=xy \pm \sqrt{x^2y^2-x^2-y^2+1}
\end{equation}
\begin{figure}[h] 
	\centering \includegraphics[scale=0.15]{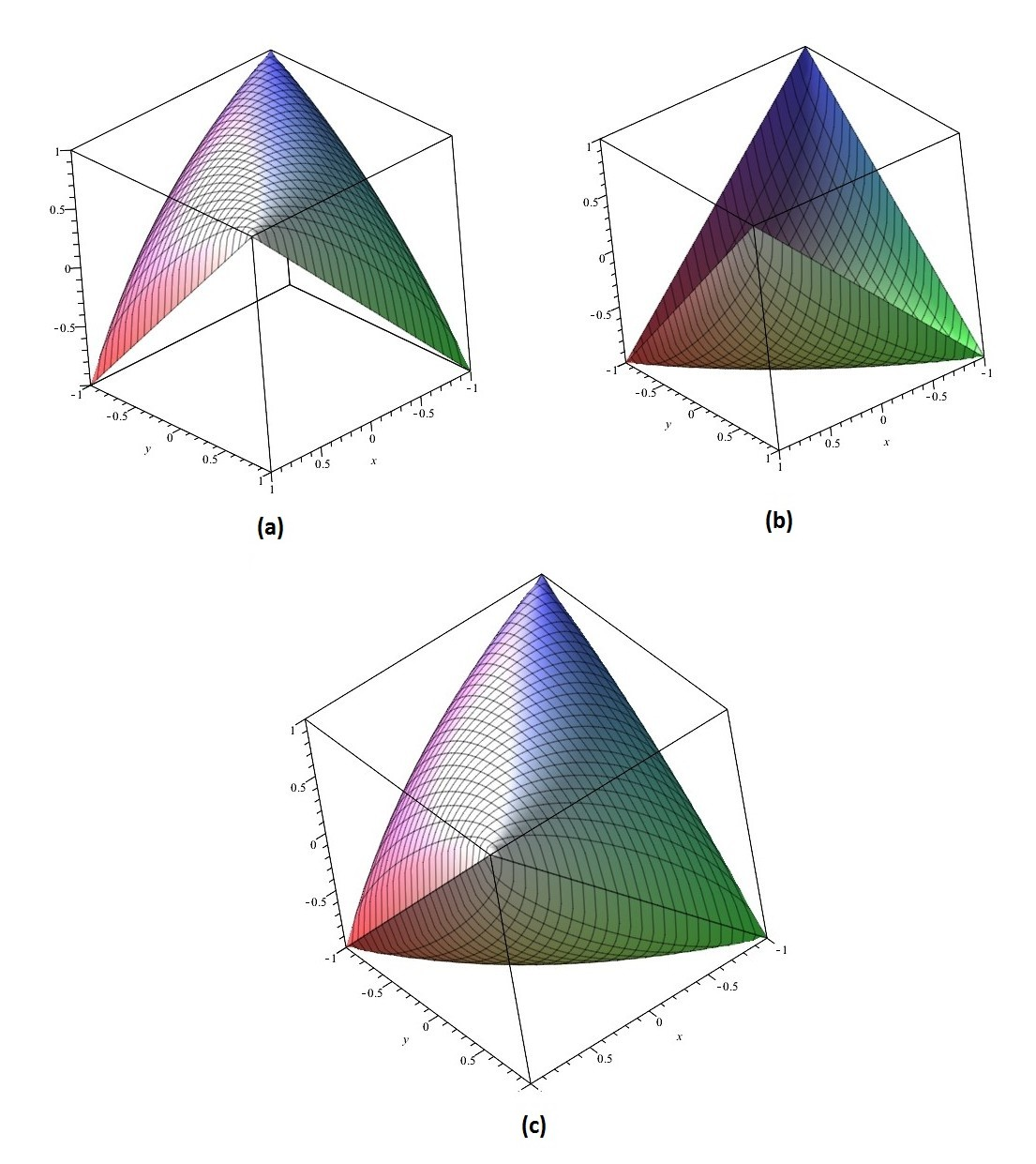} \caption{(a) Kummer superior subsurface $\Sigma_0^+$: $z^{+}=xy+\sqrt{x^2y^2-x^2-y^2+1}$, (b) Kummer inferior subsurface $\Sigma_0^-$ $: z^{-}=xy-\sqrt{x^2y^2-x^2-y^2+1}$, (c) complete Kummer subsurface $\Sigma_0$. Note that the Kummer subsurface $\Sigma_0$ is closed and its is completely inside the hypercube in this case. Thus the region between $\Sigma_0$ and the hypercube has negative $\rho$ determinant for the three assets system. } \label{nullsurface}
\end{figure}
The figure (\ref{nullsurface}) shows the Kummer superior subsurface $\Sigma_0^+$ given by $z = z^{+}(x,y)$, the Kummer inferior subsurface $\Sigma_0^-$  given by $z = z^{-}(x,y)$ and the complete Kummer subsurface $\Sigma_0$. \\ \\
Because $\Sigma_0$ separates a region with $\det \rho >0$ from that with $\det \rho <0$ and due to the origin $\vec{r}=(0,0,0)$ the determinant is one, then inside of $\Sigma_0$ the determinant of the $\rho$ matrix must be positive, which is consistent with figure 1. The region situated between $\Sigma_0$ and the cube has negative determinant in this case.  \\ \\
In terms of its diagonal form, the $\rho$ matrix inside or outside $\Sigma_0$ where $\det \rho \neq 0$, is 
\begin{equation}
\left(
\begin{matrix} \lambda_1(x,y,z) & 0 & 0 \\
0 & \lambda_2(x,y,z) & 0  \\
0 & 0 & \lambda_3(x,y,z)  \\
\end{matrix}\right) 
\end{equation}
where the three eigenvalues $\lambda_1 \ne 0$, $\lambda_2 \ne 0 $ and $\lambda_3 \ne 0$ when $\vec{r} = (x,y,z) \notin \Sigma_0$. \\ \\
On the Kummer superior subsurface $\Sigma_0^+$, the diagonal form of the $\rho$ matrix is
\begin{equation}
\left(
\begin{matrix} \lambda_1^{+}(x,y) & 0 & 0 \\
0 & \lambda_2^{+}(x,y) & 0  \\
0 & 0 & 0 \\
\end{matrix}\right) 
\end{equation}
where
\begin{equation}
\lambda_1^{+}(x,y)= \frac{3}{2}+\frac{1}{2} \sqrt{1+8x^2y^2+8xy \sqrt{x^2y^2-x^2-y^2+1}}
\end{equation}
and
\begin{equation}
\lambda_2^{+}(x,y)= \frac{3}{2}-\frac{1}{2} \sqrt{1+8x^2y^2+8xy \sqrt{x^2y^2-x^2-y^2+1}}
\end{equation}
The figure (\ref{fig2}) gives the eigenvalues $\lambda_1^{+}(x,y)$ and $\lambda_2^{+}(x,y)$ as functions of $x$ and $y$. \\
\begin{figure}[h] 
	\centering \includegraphics[scale=0.30]{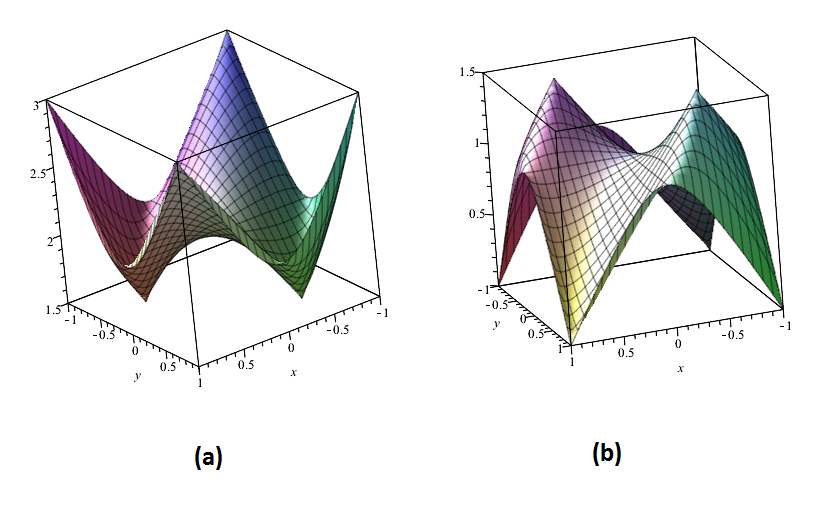} \caption{ (a) $\lambda_1^{+}(x,y)$, (b)  $\lambda_2^{+}(x,y)$} \label{fig2}
\end{figure}\\
For the Kummer inferior subsurface $\Sigma_0^-$, the diagonal form of the $\rho$ matrix is instead
\begin{equation}
\left(
\begin{matrix} \lambda_1^{-}(x,y) & 0 & 0 \\
0 & \lambda_2^{-}(x,y) & 0  \\
0 & 0 & 0 \\
\end{matrix}\right) 
\end{equation}
where
\begin{equation}
\lambda_1^{-}(x,y)= \frac{3}{2}+\frac{1}{2} \sqrt{1+8x^2y^2-8xy \sqrt{x^2y^2-x^2-y^2+1}}
\end{equation}
and
\begin{equation}
\lambda_2^{-}(x,y)= \frac{3}{2}-\frac{1}{2} \sqrt{1+8x^2y^2-8xy \sqrt{x^2y^2-x^2-y^2+1}}
\end{equation}\\
The figure (\ref{fig3}) gives the eigenvalues $\lambda_1^{-}(x,y)$ and $\lambda_2^{-}(x,y)$ as functions of $x$ and $y$. \\ \\
\begin{figure}[h] 
	\centering \includegraphics[scale=0.30]{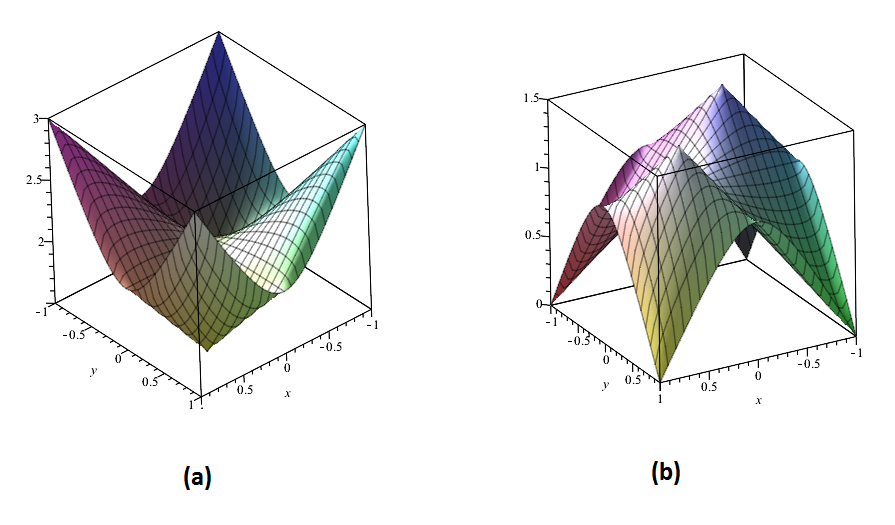} \caption{ (a) $\lambda_1^{-}(x,y)$, (b)  $\lambda_2^{-}(x,y)$} \label{fig3}
\end{figure}\\
Note that the eigenvalues $\lambda_1^{+}(x,y)$ and  $\lambda_1^{-}(x,y)$ are always greater than zero, but $\lambda_2^{+}(x,y)$ and $\lambda_2^{-}(x,y)$ are zero for the extremal values of the correlation parameter $x=\pm 1$ and $y=\pm 1$. Figure (\ref{lambda2plusminus}) shows both eigenvalues $\lambda_2^{+}(x,y)$ and $\lambda_2^{-}(x,y)$ in the same graph. One can see clearly that the $\lambda_2 (x,y)$ proper value becomes equal to zero only for the extreme correlations value cases
\begin{equation} \label{extremalvertices}
	\vec{r} = (1,1,1), \ \vec{r} = (1,-1,-1), \ \vec{r} = (-1,1,-1), \ \vec{r} = (-1,-1,1)
\end{equation}
which are the vertexes of the Kummer $\Sigma_0$ subsurface in the figure (\ref{nullsurface}) or 
the four base points of the figure (\ref{lambda2plusminus}). 
\begin{figure}[h] 
	\centering \includegraphics[scale=0.30]{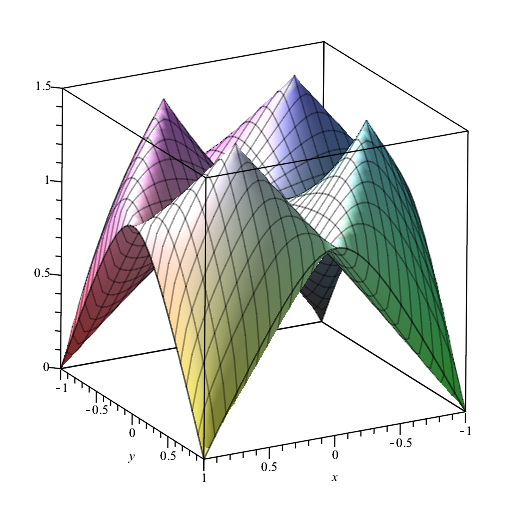} \caption{ the $\lambda_2$ eigenvalue as function of $(x,y)$} \label{lambda2plusminus}
\end{figure}\\
Thus, depending on which region of the three dimensional cube the vector $\vec{r}=(x,y,z)$ is lying, the correlation matrix $\rho$ has  two null eigenvalues, one null eigenvalue or it can be invertible. Thus the rank of the $\rho$ matrix changes when $\vec{r}$ moves along the Kummer surface. \\ \\

\section{Pricing, the Wei-Norman theorem, propagators and $\Sigma_K$}
We now tackle the problem of pricing the multi-asset option $\Pi$
by taking into account the geometrical properties of the correlation $\rho$  matrix analyzed in the section 3. In order to do that one needs first to solve the equation (\ref{difussiondiag}). For this, we apply the Wey-Norman theorem \cite{39}, \cite{40}, \cite{41}, \cite{42} that in our case this theorem establishes that the solution of (\ref{difussiondiag}) can be writing as
\begin{equation}
\Psi(\vec{\zeta},t)=U(t,0) \Psi(\vec{\zeta},0)
\end{equation}
where
\begin{equation}
U(t,0)=\Large{\Pi}_{k=1}^N \ \ e^{\biggl[ a_k(t) L_k \Biggr]}
\end{equation}
with
\begin{equation}
a_k(t)=\int_0^t \ \frac{1}{2} \lambda_k(\vec{r}) \ dt =  \frac{1}{2} \lambda_k(\vec{r}) \ t
\end{equation}
and
\begin{equation}
L_k=\frac{\partial^2}{\partial \zeta_k^2}
\end{equation}
that is
\begin{equation}
\Psi(\vec{\zeta},t)=\Large{\Pi}_{k=1}^N \ \ e^{\biggl[ \frac{1}{2} \lambda_k(\vec{r})  t \frac{\partial^2}{\partial \zeta_k^2} \Biggr]} \Psi(\vec{\zeta},0)
\end{equation}
by inserting $N$ one dimensional Dirac's deltas, one can write the above equation as
\begin{equation}
\Psi(\vec{\zeta},t)= \Large{\Pi}_{k=1}^N \ \ e^{\biggl[ \frac{1}{2} \lambda_k(\vec{r})  t \frac{\partial^2}{\partial \zeta_k^2} \Biggr]}  \biggl[ \Large{\Pi}_{m=1}^N  \int d\zeta^{'}_m \delta(\zeta_m - \zeta^{'}_m)  \biggr] \Psi(\vec{\zeta^{'}},0)
\end{equation}
or as
\begin{equation} \label{convolution}
\Psi(\vec{\zeta},t)= \int K_{\Psi}(\vec{\zeta}, t \vert \vec{\zeta}' 0) \Psi(\vec{\zeta'},0) d\vec{\zeta}' 
\end{equation} 
where the propagator $K_{\Psi}$ is defined by
\begin{equation} \label{propagatorzeta}
K_{\Psi}(\vec{\zeta}, t \vert \vec{\zeta}' 0)= \Large{\Pi}_{k=1}^N \ \ e^{\biggl[ \frac{1}{2} \lambda_k(\vec{r})  t \frac{\partial^2}{\partial \zeta_k^2} \Biggr]}  \biggl[  \delta^{(N)}(\vec{\zeta} - \vec{\zeta^{'}})  \biggr]
\end{equation} 
with $\delta^{(N)}(\vec{\zeta} - \vec{\zeta^{'}})$ the $N$ dimensional Dirac's delta. Now using the fourier expansion 
\begin{equation} \label{propagatorzeta}
 \delta^{(N)}(\vec{\zeta} - \vec{\zeta^{'}}) = \int \frac{d \vec{p}}{(2\pi)^N} \  e^{i \vec{p} \cdotp (\vec{\zeta}-\vec{\zeta^{'}} ) }
\end{equation} 
the propagator can be written finally as the product
\begin{equation} \label{propagatorproduct}
K_{\Psi}(\vec{\zeta}, t \vert \vec{\zeta}' 0) = \Large{\Pi}_{k=1}^N \Biggl[  \int \frac{d p_k}{2\pi} \ e^{ -\frac{1}{2} \lambda_k(\vec{r})  t p_k^2 +i p_k (\zeta_k - \zeta^{'}_k) }      \Biggr]
\end{equation} 
\subsection{The propagator inside $\Sigma_0$}
When $\vec{r}$ is inside of $\Sigma_0$, all eigenvalues $\lambda_k(\vec{r})$ are positive, so the $N$ integrations in (\ref{propagatorproduct}) can be performed to give \cite{45}, \cite{46}
\begin{equation} \label{propagatorzetafinal}
K_{\Psi}(\vec{\zeta}, t \vert \vec{\zeta}' 0) = \Large{\Pi}_{k=1}^N \Biggl[ \frac{1}{\sqrt{2\pi \lambda_k t}} \ e^{ -\frac{(\zeta_k - \zeta^{'}_k)^2}{2 \lambda_k t} } \Biggr]
\end{equation} 
or
\begin{equation} \label{propagatorzetafinal2}
K_{\Psi}(\vec{\zeta}, t \vert \vec{\zeta}' 0) = \frac{1}{\sqrt{(2\pi t)^N  \lambda_1 \lambda_2 \cdots \lambda_N}} \ e^{ \sum_{k=1}^{N} -\frac{(\zeta_k - \zeta^{'}_k)^2}{2 \lambda_k t} } 
\end{equation} 
By using transformations (\ref{logaS}), (\ref{PiPsi}), (\ref{chisigma}) and (\ref{forwardtime}) one can write the propagator for the option price $\Pi(\vec{S}, \tau)$ in the $(\vec{S}, \tau)$ space as 
\begin{equation} \label{propagatorregularS2}
K_{\Pi}(\vec{S}, \tau \vert \vec{S}' T) =
\frac{\exp(-r(T-\tau))}{\sqrt{(2\pi (T-\tau))^N \det(\rho)  }  \ \sigma_1 \sigma_2 \cdots \sigma_N \  S'_1 S'_2 \cdots S'_N} e^{-\bigg[ \frac{(\vec{\alpha}^t  \rho^{-1} \vec{\alpha})}{2 (T-\tau) } \bigg]}    
\end{equation}
with
\begin{equation} \label{alphai}
\vec{\alpha}_i =\vec{\alpha}_i (S_i,S'_i) = \frac{\ln(\frac{S_i}{S'_i})+(r-\frac{1}{2} \sigma_i^2)(T-\tau)}{\sigma_i}
\end{equation}
which is the usual form of the propagator in the $S$ space (see for example \cite{4}, \cite{11}). Note this form of the propagator is valid only when $\det(\rho) > 0$. So we can apply (\ref{propagatorregularS2}) inside the closed subsurface $\Sigma_0$ or some region between $\Sigma_0$ and the interior of the hypercube that verifies $\det(\rho) > 0$ and have only positive eigenvalues.

\subsection{The propagator for the Kummer surface $\Sigma_K$}
In this section we obtain an expression for the propagator over the Kummer surface $\Sigma_K$. We assume that we are in a region  $\Sigma_{K_{N_B}}$ of $\Sigma_K$ that has $N_A$ non zero eigenvalues and $N_B=N-N_A$ null eigenvalues. Due to we are on the $\Sigma_K$ surface, the equation (\ref{deternull}) implies that one of the coordinates of the $\vec{r}$ vector, is determined by the other $M-1$ coordinates. We call this independent coordinates $x_1, x_2, \cdots, x_{M-1}$. Thus in this section, the vector $\vec{r}$ is an $M$ dimensional vector that depends on $M-1$ independent coordinates. In this situation  the propagator in (\ref{propagatorproduct}) gives
\begin{equation} \label{propagatorproduct2}
K_{\Psi}(\vec{\zeta}, t \vert \vec{\zeta}' 0) = \Biggl[ \Large{\Pi}_{k=1}^{N_A}   \int \frac{d p_k}{2\pi} \ e^{ -\frac{1}{2} \lambda_k(\vec{r})  t p_k^2 +i p_k (\zeta_k - \zeta^{'}_k) }  \Biggr]      \Biggl[ \Large{\Pi}_{j=1}^{N_B}   \int \frac{d p_j}{2\pi} \ e^{i p_j (\zeta_j - \zeta^{'}_j) }      \Biggr]
\end{equation} 
By performing the integrations we arrive at
\begin{equation}  \label{propazetainsigma0}
K_{\Psi}(\vec{\zeta}, t \vert \vec{\zeta}' 0) = \frac{e^{ \sum_{k=1}^{N_A} -\frac{(\zeta_k - \zeta^{'}_k)^2}{2 \lambda_k t} }}{\sqrt{(2\pi t)^{N_A}  \lambda_1 \lambda_2 \cdots \lambda_{N_A}}} \  \Large{\Pi}_{j=1}^{N_B} \delta(\zeta_j - \zeta^{'}_j)
\end{equation} 
If we separate the $N$ dimensional vector $\vec{\zeta}$ in two parts
as
\begin{equation}
\vec{\zeta}=
\left( \begin{matrix} 
\zeta_1 \\
\vdots \\
\zeta_{N_A} \\
\zeta_{N_A+1} \\
\vdots \\
\zeta_{N} \\ 
\end{matrix}\right) = 
\left( \begin{matrix} 
\left( \begin{matrix} 
\zeta_1 \\
\vdots \\
\zeta_{N_A} \\
\end{matrix} \right) \\
 \left( \begin{matrix}
\zeta_{1} \\
\vdots \\
\zeta_{N_B} \\ 
\end{matrix} \right)
\end{matrix}\right) = 
\left( \begin{matrix} 
\vec{\zeta}_A \\
\vec{\zeta}_B
\end{matrix}\right) 
\end{equation}
the above propagator can be written in a more compact form as
\begin{equation} \label{propacompacto}
K_{\Psi}(\vec{\zeta}, t \vert \vec{\zeta}' 0) = \frac{1}{\sqrt{(2\pi t)^{N_A} \det (D_A)}} \ e^{\frac{(\vec{\zeta}_A-\vec{\zeta^{'}_A})^t D^{-1}_A (\vec{\zeta}_A-\vec{\zeta^{'}_A})}{2 t}} \ \delta^{(N_B)} (\vec{\zeta}_B-\vec{\zeta^{'}}_B)
\end{equation} 
where
\begin{equation}
D_A = \left( \begin{matrix} 
\lambda_1 & 0 & \cdots  & 0 \\
0 & \lambda_2 & \cdots & 0 \\
\vdots &  \vdots & \cdots &  \vdots \\
0 & 0 & \cdots & \lambda_{N_A} \\
\end{matrix}\right) 
\end{equation}
is the reduced diagonal $\rho$ matrix on the Kummer surface $\Sigma_K$. If one separates the vector $\vec{\chi}$ in $A$ and $B$ components as
\begin{equation}
\vec{\chi}=
\left( \begin{matrix} 
\vec{\chi}_A \\
\vec{\chi}_B
\end{matrix}\right) 
\end{equation}
then relation (\ref{zetachi}) induces the transformation
\begin{equation} \label{zetaABchiAB}
    \left(\begin{matrix} 
	\vec{\zeta}_A\\
	\vec{\zeta}_B\\
	\end{matrix}\right) \\ =
		\left(\begin{matrix} 
		U^{-1}_{AA} &  U^{-1}_{AB} \\
		U^{-1}_{BA} &  U^{-1}_{BB} \\
		\end{matrix}\right)
	 \left(\begin{matrix} 
	 	\vec{\chi}_A\\
	 	\vec{\chi}_B\\
	 \end{matrix}\right) 
\end{equation}
where $	U^{-1}_{AA}$, $	U^{-1}_{AB}$, $	U^{-1}_{BA}$ and $	U^{-1}_{BB}$ are the matrices that result from sectioning $U^{-1}$ into $A$ and $B$ components. \\ \\
The quadratic term in the exponential of (\ref{propazetainsigma0}) can be expressed in the $\chi_A$ and $\chi_B$ components as
\begin{equation} \label{cuadraticterm}
\begin{matrix} 
(\vec{\zeta}_A-\vec{\zeta'}_A)^t D^{-1}_A (\vec{\zeta}_A-\vec{\zeta'}_A) = & (\vec{\chi}_A-\vec{\chi'}_A)^t \ [U^{-1}_{AA}]^t D^{-1}_A U^{-1}_{AA} \ (\vec{\chi}_A-\vec{\chi'}_A) \ +  & \\ 
 & (\vec{\chi}_B-\vec{\chi'}_B)^t \ [U^{-1}_{AB}]^t D^{-1}_A U^{-1}_{AA} \ (\vec{\chi}_A-\vec{\chi'}_A) \ + & \\
 & (\vec{\chi}_A-\vec{\chi'}_A)^t \ [U^{-1}_{AA}]^t D^{-1}_A U^{-1}_{AB} \ (\vec{\chi}_B-\vec{\chi'}_B) \ + & \\
 & (\vec{\chi}_B-\vec{\chi'}_B)^t \ [U^{-1}_{AB}]^t D^{-1}_A U^{-1}_{AB} \ (\vec{\chi}_B-\vec{\chi'}_B) \ + & 
 \end{matrix}
\end{equation}
Now, from (\ref{zetaABchiAB}) we have
\begin{equation} \label{zetaBchiAB}
(\vec{\zeta}_B-\vec{\zeta'}_B)= U^{-1}_{BA} (\vec{\chi}_A-\vec{\chi'}_A)+U^{-1}_{BB} (\vec{\chi}_B-\vec{\chi'}_B)
\end{equation}
The Dirac's delta in (\ref{propacompacto}) implies that
\begin{equation}
0 = U^{-1}_{BA} (\vec{\chi}_A-\vec{\chi'}_A)+U^{-1}_{BB} (\vec{\chi}_B-\vec{\chi'}_B)
\end{equation}
The above equation permits writing the vector $(\vec{\chi}_B-\vec{\chi'}_B)$ in terms of $(\vec{\chi}_A-\vec{\chi'}_A)$ as
\begin{equation} \label{gamma}
(\vec{\chi}_B-\vec{\chi'}_B) = - U_{BB} U^{-1}_{BA} (\vec{\chi}_A-\vec{\chi'}_A)
\end{equation}
replacing in (\ref{cuadraticterm}) one can write the quadratic term as
\begin{equation} \label{cuadraticterm2}
(\vec{\zeta}_A-\vec{\zeta'}_A)^t D^{-1}_A (\vec{\zeta}_A-\vec{\zeta'}_A) =  (\vec{\chi}_A-\vec{\chi'}_A)^t \ \rho^{-1}_{\Sigma_K} \ (\vec{\chi}_A-\vec{\chi'}_A) 
\end{equation}
where $\rho^{-1}_{\Sigma_K}$ is defined by
\begin{equation} \label{rhoonsigmainverse}
\begin{matrix} 
\rho^{-1}_{\Sigma_K} = & [U^{-1}_{AA}]^t D^{-1}_A U^{-1}_{AA} +& \\ \\ 
& [U^{-1}_{AB} U_{BB} U^{-1}_{BA}]^t D^{-1}_A  U^{-1}_{AA} +& \\ \\
& [U^{-1}_{AA}]^t D^{-1}_A U^{-1}_{AB} U_{BB} U^{-1}_{BA} +& \\ \\
& [U^{-1}_{AB} U_{BB} U^{-1}_{BA}]^t D^{-1}_A U^{-1}_{AB} U_{BB} U^{-1}_{BA} & \\
\end{matrix}
\end{equation}
From (\ref{zetaABchiAB}) we notice that
\begin{equation}
d \vec{\zeta} = d \vec{\zeta_A} d \vec{\zeta_B} = d \vec{\chi_A} d \vec{\chi_B} =d \vec{\chi}
\end{equation}
Using (\ref{zetaBchiAB}) and (\ref{cuadraticterm2}) in  (\ref{convolution}), the option price can be written as
\begin{equation} \label{pricingchiAB}
\begin{matrix}
\Psi(\vec{\chi}_A, \vec{\chi}_B ,t)= \\
\int
\frac{e^{\frac{(\vec{\chi}_A-\vec{\chi'}_A)^t \ \rho^{-1}_{\Sigma_K} \ (\vec{\chi}_A-\vec{\chi'}_A)}{2 t}}}{\sqrt{(2\pi t)^{N_A} \det (D_A)}} \  \ \delta^{(N_B)} (U^{-1}_{BA} (\vec{\chi}_A-\vec{\chi'}_A)+U^{-1}_{BB} (\vec{\chi}_B-\vec{\chi'}_B)) \Psi(\vec{\chi'}_A,\vec{\chi'}_B, 0) \ \ d\vec{\chi'}_A  \ d\vec{\chi'}_B \\
\end{matrix}
\end{equation} 
Integrating over $d\vec{\chi'}_B$ gives
\begin{equation} \label{pricingchiAB2}
\Psi_{(\vec{r})}(\vec{\chi}_A, \vec{\chi}_B ,t)= \\
\int
\frac{e^{\frac{(\vec{\chi}_A-\vec{\chi'}_A)^t \ \rho^{-1}_{\Sigma_K} \ (\vec{\chi}_A-\vec{\chi'}_A)}{2 t}}}{\sqrt{(2\pi t)^{N_A} \det (D_A)}} \
\frac{1}{\det(U^{-1}_{BB})} \
\Psi(\vec{\chi'}_A,\vec{\chi'}_B, 0) \ \ d\vec{\chi'}_A \\
\end{equation} 
where $\vec{\chi'}_B$ must be evaluated from (\ref{gamma}) in terms of $\vec{\chi}_B$ and $(\vec{\chi}_A-\vec{\chi'}_A)$ as
\begin{equation}
\vec{\chi'}_B= \vec{\chi}_B + \gamma (\vec{\chi}_A-\vec{\chi'}_A)
\end{equation}
where the rectangular $N_B \times N_A$ matrix $\gamma$ is defined by
\begin{equation}
\gamma =  U_{BB} U^{-1}_{BA}
\end{equation}
It must be noted that $U^{-1}$, the eigenvalues $\lambda_i$, and  the rectangular matrix $\gamma$ are functions of the vector $\vec{r}$ that lies on the null surface $\Sigma_K$. Thus the option price is also a function of $\vec{r}$. Using (\ref{logaS}), (\ref{PiPsi}), (\ref{chisigma}) and (\ref{forwardtime}) one can write the option price in the $(\vec{S, \tau})$ space as $\Pi_{(\vec{r})}(\vec{S}, \tau)$ and is given by 
\begin{equation} \label{pricingSAB}
\Pi_{(\vec{r})}(\vec{S}_A, \vec{S}_B ,\tau)= \\
\int
\frac{e^{\frac{(\vec{\alpha}_A)^t  \rho^{-1}_{\Sigma_K}  (\vec{\alpha}_A)}{2 (T-\tau) }}}{\sqrt{(2\pi (T-\tau))^{N_A} \det (D_A)}} \
\frac{ e^{-r(T-\tau)}}{\det(U^{-1}_{BB})} \
\frac{\Psi(\vec{S'}_A,\vec{S'}_B, T)}{\sigma_1 \cdots \sigma_{N_A}}
 \ \ \frac{ d\vec{S'}_A}{S'_1 \cdots S'_{N_A} } \\
\end{equation} 
where the components of the $\vec{\alpha}$ are given by
\begin{equation}
\alpha_A^{ \ j}= \frac{\ln(\frac{S^{ \ j}_A}{S^{' \ j}_A})+(r-\frac{1}{2} \sigma_j^2)(T-\tau)}{\sigma_j} \ \ \ \ \ \ j=1, \cdots, N_A
\end{equation}
and the components of the vector $\vec{S'}_B$ are given in terms of $\vec{S}_A$,  $\vec{S'}_A$ and  $\vec{S}_B$ according to
\begin{equation}
S_B^{' \ i}= S_B^{\ i} \ \biggl[ \Pi_{j=1}^{N_A} \biggl(\frac{S^{ \ j}_A}{S^{' \ j}_A} \biggr)^{\frac{\sigma_i}{\sigma_j} \gamma_{ij}} \biggr] \ e^{\bigl[ (r-\frac{1}{2} \sigma_i^2) + \sum_{j=1}^{N_A}  \frac{\sigma_i}{\sigma_j} \gamma_{ij}  (r-\frac{1}{2} \sigma_j^2)  \bigr] (T-\tau)}  \ \ \ \ i=1, \cdots, N_B
\end{equation}
with $\gamma_{ij}$ the components of the rectangular matrix $\gamma$
\begin{equation}
\gamma_{ij}= [U_{BB} U^{-1}_{BA}]_{ij} \ \ \ \ \ i=1, \cdots, N_B, \ \ j=1, \cdots, N_A
\end{equation}
When $\vec{r}$ moves over the Kummer   surface $\Sigma_K$, the rank of the $\rho$ matrix can change, so the dimensions of $N_A$ and $N_B=N-N_A$ also change, but equation (\ref{pricingSAB}) is always valid.

\subsection{The propagator outside $\Sigma_0$}
When the vector $\vec{r}$ is lying outside the Kummer subsurface $\Sigma_0$, there are regions where the determinant of the correlation matrix is negative. This implies that the propagator given in (\ref{propagatorregularS2}) becomes complex.
But, worse than that, in this case one of the eigenvalues $\lambda_k$ is negative, so the propagator given in (\ref{propagatorzetafinal2}) generates an exponential growth in the associated $\zeta_k$ coordinate. Then the convolution in (\ref{convolution}) is not well defined. Thus, one cannot price the option in regions outside the Kummer subsurface $\Sigma_0$ that have negative $\rho$ determinant.

\section{Conclusions and further research}

In this research, we have analyzed in detail the existence of the 
solution of the multi-asset Black-Scholes model. We point out 
that the correlation parameter space, which is equivalent to an $N$
dimensional hypercube, limite the existence of a valid solution 
for the multi-asset Black-Scholes model. Particularly, we show 
that inside of this hypercube there is a surface, called the Kummer surface $\Sigma_{K}$,  where the determinant of the correlation matrix $\rho$ is zero, so the usual formula for the  propagator of the $N$ asset Black-Scholes equation is no longer valid. We also study in detail the case for three assets and its implied geometry when the determinant of the correlation matrix is zero. Finally, by using the Wei-Norman theorem, we compute the propagator over the variable rank surface $\Sigma_{K}$ for the general $N$ asset case, which is applicable over all the Kummer surface, whatever be the rank of the $\rho$ matrix. This formulation corrects the past solution of this problem and its extensions. \\
As future research, most of the papers related to the multi-asset Black-Scholes model must be revisited in line of our results, as well as others where it is implicitly assumed that a well behaved multivariate Gaussian distribution must exist, as is the case of the stochastic volatility family (see for instance \cite{47},  \cite{48}).

\begingroup\raggedright

\endgroup
\end{document}